# Low intensity threshold values for keV X-ray production in laser-cluster interaction[a]

C. Prigent[b], L. Adoui[c], E. Lamour, J.-P. Rozet[d], D. Vernhet

INSP, CNRS UMR 7588, Universités P. et M. Curie and D. Diderot,

140 rue de Lourmel, 75015 Paris, France

O. Gobert, P. Meynadier, D. Normand and M. Perdrix

CEA-DSM/DRECAM/SPAM, C.E. Saclay, Bat. 522, 91191 Gif-sur-Yvette Cedex, France

C. Deiss, N. Rohringer, and J. Burgdörfer

Institute for Theoretical Physics, Vienna University of Technology, A-1040 Vienna, Austria, EU

(Dated : July 4, 2005)

Absolute Ar $X_K$ (3 keV) and Xe $X_L$ (4 keV) photon yields are measured as a function of laser peak intensity, for large ($10^4$ - $10^5$ atoms) argon and xenon clusters submitted to intense laser pulses. A clear threshold behavior is observed. The intensity threshold is found as low as $2 \; 10^{14}$ Wcm$^{-2}$, a value matching the production of singly-charged argon or doubly-charged xenon by optical field ionization. Possible mechanisms involved in the inner-shell ionization are discussed. The outcome on the optimization of keV X-ray production is underlined.

---

[a] Experiment performed at LUCA-SPAM C.E. Saclay

[b] Present address: Max-Planck-Institut für Kernphysik, D-69029 Heidelberg, Germany

[c] Current address: CIRIL, Rue C. Bloch BP 5133 14070 Caen Cedex 5, France

[d] Corresponding author. Electronic address: rozet@gps.jussieu.fr

A fascinating feature of intense laser-matter interaction is its efficiency for converting photons in the eV range to X-rays in the keV. It is now well known that large clusters, similarly to solids, couple very efficiently with intense subpicosecond laser pulses. A highly ionized and excited medium is formed at the cluster scale (a few ten nanometers), and treated in some models [1,2,3] as a "nanoplasma". Highly charged ions with energies reaching the MeV [4,5], electrons with energies up to a few keV [6], and UV [2] and keV X-ray photons from inner-shell ionized atoms are emitted [7,8,9,10].

X-ray spectroscopy allows to perform quantitative measurements, and short lifetime states down to a few ten femtoseconds are observed. This gives access to the dynamical evolution of the irradiated cluster on a time scale comparable to the laser pulse duration, and much smaller than reached in the case of ion spectroscopy. It must be noted, however, that only ions with inner-shell vacancies can be observed by X-ray spectroscopy. Up to now, X-ray emission studies have nevertheless been mainly limited to qualitative observations. A systematic study of absolute X-ray emission yields and ionic charge state distributions as a function of various physical parameters governing the laser-cluster interaction (cluster size, atomic number and total atomic density on one hand, laser pulse energy, pulse duration, wavelength and polarization on the other hand) was still lacking. In this Letter, we present the results of such a quantitative study, here for X-ray yields as a function of laser peak intensity. New possible mechanisms for the production of highly charged ions with *inner-shell* vacancies are considered.

After a decade of experimental and theoretical studies devoted to these phenomena, most of the models which have been developed so far still remain controversial[11]. In all models some electrons are first released from atoms in the cluster by optical field ionization (OFI). However, this process alone cannot explain the production of, for instance, highly ionized argon ions with a K vacancy, since a minimum intensity of $4\ 10^{21}$ Wcm$^{-2}$ would be necessary in the barrier suppression ionization approximation. Two main mechanisms have been proposed to account for the production of highly charged ions. Further ionization can take place through electron-ion collisions (inelastic scattering) in the cluster as assumed in the nanoplasma model [2], where fast electrons are produced by inverse bremsstrahlung. This process becomes very efficient when the plasma frequency in the cluster matches the laser frequency. Megi et al. [12] have performed calculations showing, however, that this resonance should be strongly damped, and predict much lower electron temperatures. Alternatively, ionization may be assisted by an increase of the electric field inside the cluster due to the combined charge of electrons and ions, as introduced by Rose-Petruck et al. [13] in the "Ignition Ionization Model". None of these models leads to the prediction of hard X-ray emission. In this context, we have proposed a simple Inner-Shell Ionization (ISI) model [9] where ionization is due to the inelastic collisions of electrons moving freely in the laser field. This model uses the concept of *effective focal volume* corresponding to the region of space where the laser intensity is large enough to induce an electron velocity sufficient to produce K-shell or L-shell vacancies. A prediction of this model is an intensity threshold value of a few $10^{16}$ Wcm$^{-2}$ for Xe L-shell or Ar K-shell ionization. The data presented in the following show, in fact, intensity thresholds one or two orders of magnitude smaller.

The experimental setup, including the apparatus used for cluster generation, has been described in detail in previous publications [8,9]. Briefly, clusters are generated within a pulsed adiabatic expansion of the well documented Hagena type [14,15]. This device leads to a mean atomic density proportional to the backing pressure ($P_0$), and a mean cluster size scaling [16,17] as $P_0^{1.8}$. More details will be given in a forthcoming paper devoted to the evolution of X-ray yields with cluster size. The intense laser field is generated with a Ti:sapphire laser system delivering pulses of 50 fs minimum duration at 800 nm with a repetition rate of 20 Hz. The beam diameter is approximately 50 mm and the maximum pulse energy available in the interaction zone is 100 mJ. The laser light is focused with a f = 480 mm lens leading to a $1/e^2$ beam waist of ~ 30 μm and maximum peak intensities $I_{peak}$ ~ $10^{17}$ Wcm$^{-2}$. The $I_{peak}$ values are determined by imaging the focal point on a CCD camera and from systematic measurements via autocorrelation techniques of the laser pulse duration taking into account the temporal aberrations induced by the lens. Total uncertainties on $I_{peak}$ values are estimated to be ~ 20%.

X-rays are analyzed using two Si(Li) detectors and a high-resolution high-transmission Bragg-crystal spectrometer. A description of these spectrometers[18] and of the procedure used to determine absolute yields [8,9] has been given elsewhere. In the case of Ar clusters, the X-ray spectrum observed with a Si(Li) detector, exhibits as a main feature a Kα transition at 3.1 keV, with a low energy background due to bremsstrahlung. With the crystal spectrometer, several components in the Kα are resolved (as shown in Fig. 1), which correspond to different charge states between 12+ up to 16+ (heliumlike). For such a typical spectrum, recorded in the regime of high laser intensity, the mean charge state is close to 14.5. Absolute X-ray yields determined independently from Si(Li) detectors and crystal spectrometer, are found in good agreement. Uncertainties on these yields are in the range between 15 to 30%.

Fig. 2 presents the X-ray yield dependence on laser peak intensity in the case of argon clusters and three laser pulse durations (60, 150 and 610 fs). Similar results have been obtained with Xe clusters. All recorded data show the same behavior: a rapid increase of the X-ray yield above a well defined threshold is followed by a slower ramp accurately described by a $I_{peak}^{3/2}$ law. In all cases, the X-ray yields can be precisely fitted by the expansion of the *effective focal volume* with the laser intensity. This effective focal volume, where, for a particular laser peak intensity ($I_{peak}$), the laser intensity exceeds a given threshold intensity ($I_{th}$), is given by [19]

$$V_{eff.foc.} = \frac{(\pi w_0^2)^2}{\lambda} \left\{ \frac{4}{3} \cdot \left( \frac{I_{peak}}{I_{th}} - 1 \right)^{1/2} + \frac{2}{9} \cdot \left( \frac{I_{peak}}{I_{th}} - 1 \right)^{3/2} - \frac{4}{3} \cdot arctg\left( \frac{I_{peak}}{I_{th}} - 1 \right)^{1/2} \right\}$$

with $w_0$, the laser beam waist. The fact that the X-ray yield follows this law suggests a *constant* (or nearly constant) X-ray emission probability inside this volume, as soon as an intensity threshold is achieved. This is in agreement with the assumptions of the ISI model. Simple calculations show in fact that our measurements can also be reproduced if this probability slightly increases with laser intensity above the threshold, but not faster than linearly [17]. For a given intensity, we can calculate the number of atoms in the effective focal volume deduced from the measured threshold intensities. The absolute X-ray yield divided by the number of atoms gives the X-ray emission probability per atom. For intensities exceeding 2 or 3 times the threshold value (saturation regime), the mean charge state deduced from high resolution spectra is constant. Therefore, assuming a mean fluorescence yield ~ 0.35 for multiply ionized ions [20], inner-shell *ionization probabilities* per atom can be estimated. Table I gives the measured threshold values for argon and xenon clusters, and the inferred ionization probabilities. Though more systematic studies are needed, a comparison of threshold values for two Ar cluster sizes differing by a factor ~ 5 has also been performed, and exhibits a very weak dependence on this parameter. However, the threshold values appear to strongly depend on the laser pulse duration. A fast decrease, between 60 fs and ~ 300 fs, is followed by an almost constant value, up to 2 ps, of ~ 2 - 3 $10^{14}$ Wcm$^{-2}$ (see Fig. 3). From the available data, it appears that below ~ 300 fs, the threshold *fluence* (or energy) remains almost constant. Clearly more measurements are nevertheless needed to determine accurately this behavior.

The most remarkable result obtained here is the unexpectedly low threshold values, up to two orders of magnitude smaller than what was predicted by the ISI model [9]. Indeed, the maximum free electron oscillation energy in the laser field (2.$U_p$) may be as low as ~ 30 eV at threshold (see Table I). This is barely enough to produce Ar$^{1+}$ and Xe$^{2+}$ either by inelastic electron-ion collisions, or even by OFI. Similarly, the electron temperatures which can be deduced from the nanoplasma model [12] are far too small to explain inner-shell ionization (typical maximum electron temperatures are ~ 25 eV for a 2.8 $10^5$ atoms cluster and a 60 fs laser pulse of 2 $10^{15}$ Wcm$^{-2}$). Alternative explanations have to be considered for the electron-heating process up to keV energies. In recent experiments an anisotropic emission of fast ions has been observed [21,22]. This feature has been explained by calculations [23,24] including charge polarization of the cluster and field enhancement inside the cluster. This charge polarization effect follows the line of the Ignition Ionization Mechanism introduced by Rose-Petruck et al.[13]. Unfortunately, quantitative predictions have been mainly limited so far to small ($\leq 1000$ atoms) clusters.

One mechanism which could account for the efficient heating of the electrons well beyond the ponderomotive energy $U_p$, is acceleration in the laser field by simultaneous elastic scattering at the core potential of cluster ions[25]. Elastic backscattering at the atomic potential of the ions can flip the velocity vector of an electron, allowing it, with non-negligible probability, to remain synchronized with the alternating laser field vector. The electron will thus efficiently accumulate, rather than lose, momentum during subsequent laser half-cycles, thereby rapidly gaining kinetic energy well beyond the ponderomotive limit. This heating mechanism is closely connected to the Fermi shuttle acceleration[26] and also to the lucky-electron model proposed for IR photoemission from metallic surfaces[27]. A realistic estimate for the efficiency of elastic electron-ion scattering as a heating mechanism critically depends on a correct determination of the differential scattering cross-section for large scattering angles. The latter is governed by the short-ranged non-Coulombic core potential leading to the dominance of few low- order partial waves and to Ramsauer-Townsend interference oscillations. Approximations, such as the widely used softened Coulomb potential[2,24], greatly underestimate the probability for backscattering. First results of simulations using scattering cross-sections determined by parametrized Hartree-Fock potentials for various charge states of cluster ions and a Classical Trajectory Monte Carlo simulation for the electron dynamics inside the cluster dynamics[28] for large argon clusters ($2.8 \times 10^5$ atoms) show the onset of x-ray emission for short pulses already in the low- laser intensity region ($10^{15}$-$10^{16}$ Wcm$^{-2}$).

Another important result is relevant to the X-ray yield optimization. Since the X-ray emission probability above the threshold appears to be constant (or almost constant), the total X-ray yield is strictly proportional to the effective focal volume. Depending on the threshold value and the maximum pulse energy available, the X-ray yield may be increased by increasing the beam waist, in contradiction to what could be anticipated from available models. Fig. 4, where the effective volume is plotted as a function of beam waist values, illustrates this behavior.

In summary, studies of the X-ray yield dependence on laser peak intensity have demonstrated a clear threshold behavior, suggesting a constant inner-shell ionization probability in the saturation regime. Unexpectedly low threshold values, as low as $2\ 10^{14}$ Wcm$^{-2}$, have been measured. This indicates that ignition of inner-shell ionization immediately follows the production of the first ionic species. None of the previously proposed theoretical models provides a satisfactory explanation for an high electron temperature at such moderate intensities. Large-angle backscattering at ions, resembling the Fermi shuttle, is proposed as a new efficient electron heating process. Further experimental investigations of the X-ray yield dependence on the laser wavelength, pulse duration as well as on the cluster size are in progress. In parallel, more complete tests of the dynamical simulations as a function of these parameters are planed.

Work is supported by FWF SFB-16 (Austria).

[1] T. Ditmire, T. Donnelly, R.W. Falcone, and M.D. Perry, Phys. Rev. Lett. **75**, 3122 (1995).

[2] T. Ditmire, T. Donnelly, A.M. Rubenchik, R.W. Falcone, and M.D. Perry, Phys. Rev. A **53**, 3379 (1996).

[3] M. Schmidt, and D. Normand, Physics World **10**, 26 (1997).

[4] T. Ditmire, J.W.G. Tisch, E. Springate, M.B. Mason, N. Hay, R.A. Smith, J. Marangos, and M.H.R. Hutchinson, Nature (London) **386**, 54 (1997).

[5] M. Lezius, S. Dobosz, D. Normand, and M. Schmidt, J. Phys. B: At. Mol. Opt. Phys. **30**, L251 (1997).

[6] Y.L. Shao, T. Ditmire, J.W.G. Tisch, E. Springate, J.P. Marangos, and M.H.R. Hutchinson, Phys. Rev. Lett. **77**, 3343 (1996).

[7] A. McPherson, B.D. Thompson, A.B. Borisov, K. Boyer, and C.K. Rhodes, Nature (London) **370**, 631 (1994).

[8] S. Dobosz, M. Lezius, M. Schmidt, P. Meynadier, M. Perdrix, D. Normand, J.-P. Rozet, and D. Vernhet, Phys. Rev. A **56**, R2526 (1997).

[9] J.-P. Rozet, M. Cornille, S. Dobosz, J. Dubau, J.-C. Gauthier, S. Jacquemot, E. Lamour, M. Lezius, D. Normand, M. Schmidt, and D. Vernhet, Physica Scripta **T92**, 113 (2001).

[10] V. Kumarappan, M. Krishnamurthy, D. Mathur, and L.C. Tribedi, Phys. Rev. A **63**, 023203 (2001).

[11] V.P. Krainov, and M.B. Smirnov, Phys. Rep. **370**, 237 (2002).

[12] F. Megi, M. Belkacem, M.A. Bouchene, E. Suraud, and G. Zwicknagel, J. Phys. B: At. Mol. Opt. Phys. **36**, 273 (2003).

[13] C. Rose-Petruck, K.J. Schafer, K.R. Wilson, and C.P.J. Barty, Phys. Rev. A **55**, 1182 (1997).

[14] O.F. Hagena, and W. Obert, J. Chem. Phys. **56**, 1793 (1972).

[15] O.F. Hagena, Z. Phys. D : Atoms, Molecules and Clusters **4**, 291 (1987).

[16] F. Dorchies, F. Blasco, T. Caillaud, J. Stevefelt, C. Stenz, A.S. Boldarev, and V.A. Gasilov, Phys. Rev. A **68**, 023201 (2003).

[17] C. Prigent, PhD thesis, Paris (2004), http://tel.ccsd.cnrs.fr.

[18] D. Vernhet, J.P. Rozet, I. Bailly-Despiney, C. Stephan, A. Cassimi, J.P. Grandin, and L.J. Dubé, J. Phys. B: At. Mol. Opt. Phys. **31**, 117 (1998).

[19] S. Augst, D. Meyerhofer, D. Strickland, and S.L. Chin, J. Opt. Soc. Am. B **8**, 858 (1991).

[20] C.P. Bhalla, Phys. Rev. A **8**, 2877 (1973).

[21] E. Springate, N. Hay, J.W.G. Tisch, M.B. Mason, T. Ditmire, M.H.R. Hutchinson, and J.P. Marangos, Phys. Rev. A **61**, 063201 (2000).

[22] V. Kumarappan, M. Krishnamurthy, and D. Mathur, Phys. Rev. Lett. **87**, 085005 (2001).

[23] K. Ishikawa and T. Blenski, Phys. Rev. A **62**, 063204 (2000).

[24] C. Jungreuthmayer, M. Geissler, J. Zanghellini, and T. Brabec, Phys. Rev. Lett. **92**, 133401 (2004).

[25] C. Deiss, N. Rohringer, and J. Burgdörfer, Nucl. Instrum. Methods Phys. Res. B **235**, 210 (2005).

[26] E. Fermi, Phys. Rev. **75**, 1169 (1949).

[27] F. Pisani, J.L. Fabre, S. Guizard, P. Palianov, Ph. Martin, F. Glotin, and J.M. Ortega, Phys. Rev. Lett. **87**, 187403 (2001).

[28] C. Deiss et al. to be published

**TABLES**

**TABLE I. Threshold values (laser intensity and electric field), maximum oscillation electron energies (2.Up) and ionization probabilities for argon and xenon clusters for various laser pulse duration and cluster type.**

| Cluster type | Backing pressure (bar) | Mean cluster size (atoms) | Atomic density ($cm^{-3}$) | Pulse duration (fs) | Threshold intensity ($Wcm^{-2}$) | Threshold laser field (a.u.) | 2.Up (eV) | Ionization probability |
|---|---|---|---|---|---|---|---|---|
| Ar | 13 | $3.7\ 10^4$ | $9.4\ 10^{16}$ | 60 | $2.9\ 10^{15}$ | 0.267 | 345 | ~ $2.\ 10^{-6}$ |
| Ar | 13 | $3.7\ 10^4$ | $9.4\ 10^{16}$ | 150 | $1.2\ 10^{15}$ | 0.185 | 145 | ~ $2.\ 10^{-6}$ |
| Ar | 13 | $3.7\ 10^4$ | $9.4\ 10^{16}$ | 610 | $2.8\ 10^{14}$ | 0.089 | 35 | ~ $2.\ 10^{-6}$ |
| Xe | 9 | $1.7\ 10^5$ | $6.5\ 10^{16}$ | 60 | $3.5\ 10^{15}$ | 0.315 | 420 | ~ $1.5\ 10^{-4}$ |
| Xe | 9 | $1.7\ 10^5$ | $6.5\ 10^{16}$ | 320 | $2.5\ 10^{14}$ | 0.084 | 30 | ~ $6.\ 10^{-4}$ |
| Xe | 3 | $2.8\ 10^4$ | $2.4\ 10^{16}$ | 2360 | $2.2\ 10^{14}$ | 0.079 | 26 | ~ $1.\ 10^{-4}$ |

FIGURE CAPTIONS

FIG. 1. Ar 1s2pnl → 1s$^2$nl high-resolution spectrum recorded during irradiation of argon clusters by 60 fs pulses at 800 nm with a peak intensity of 3 10$^{16}$ Wcm$^{-2}$ and a backing pressure of 20 bars.

FIG. 2. X-ray yield variation with laser peak intensity obtained during irradiation of argon clusters by 60, 150 and 610 fs pulses at 800 nm with a backing pressure of 13 bars. Lines correspond to the calculated effective focal volume variation with laser intensity (see text).

FIG. 3. Threshold intensities as a function of laser pulse duration for Ar and Xe clusters. Open (Ar) and closed (Xe) symbols are experimental results. Solid and dotted lines correspond to the OFI threshold for producing Ar$^{1+}$ and Xe$^{2+}$ respectively. Dashed line is plotted to guide the eyes.

FIG. 4. Evolution of the effective focal volume with the beam waist for several laser tenergies. The threshold peak intensity and pulse duration are assumed to be 2. 10$^{15}$ Wcm$^{-2}$ and 50 fs respectively.

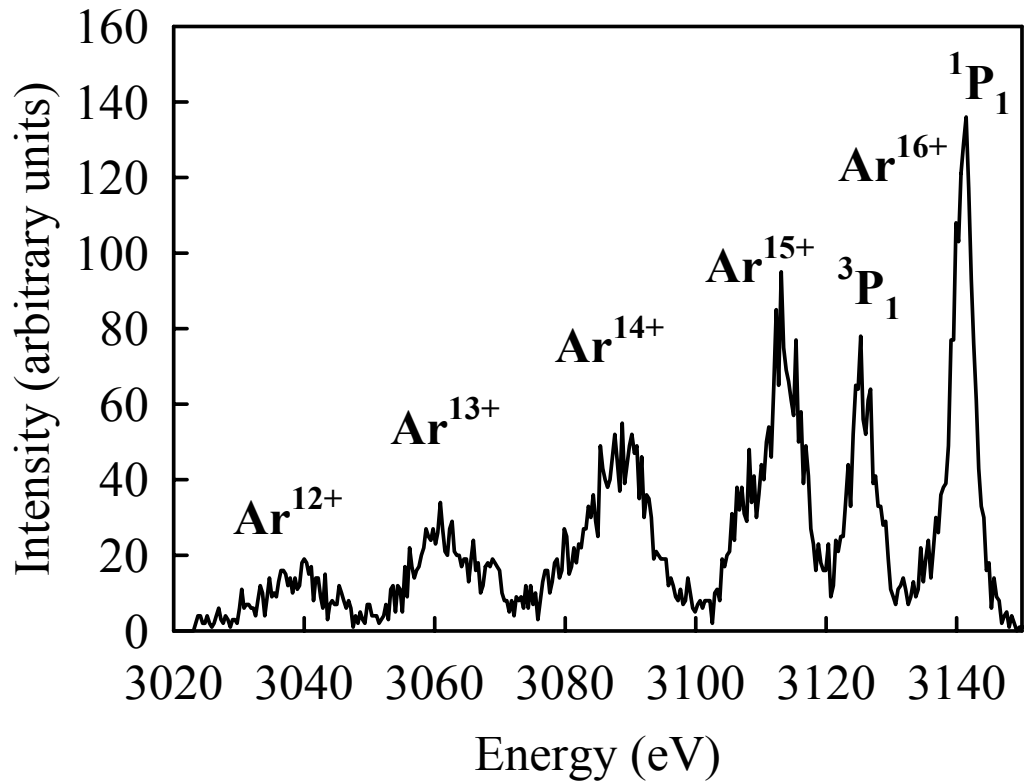

FIG. 1. Ar 1s2pnl → 1s²nl high-resolution spectrum recorded during irradiation of argon clusters by 60 fs pulses at 800 nm with a peak intensity of 3 10¹⁶ Wcm⁻² and a backing pressure of 20 bars.

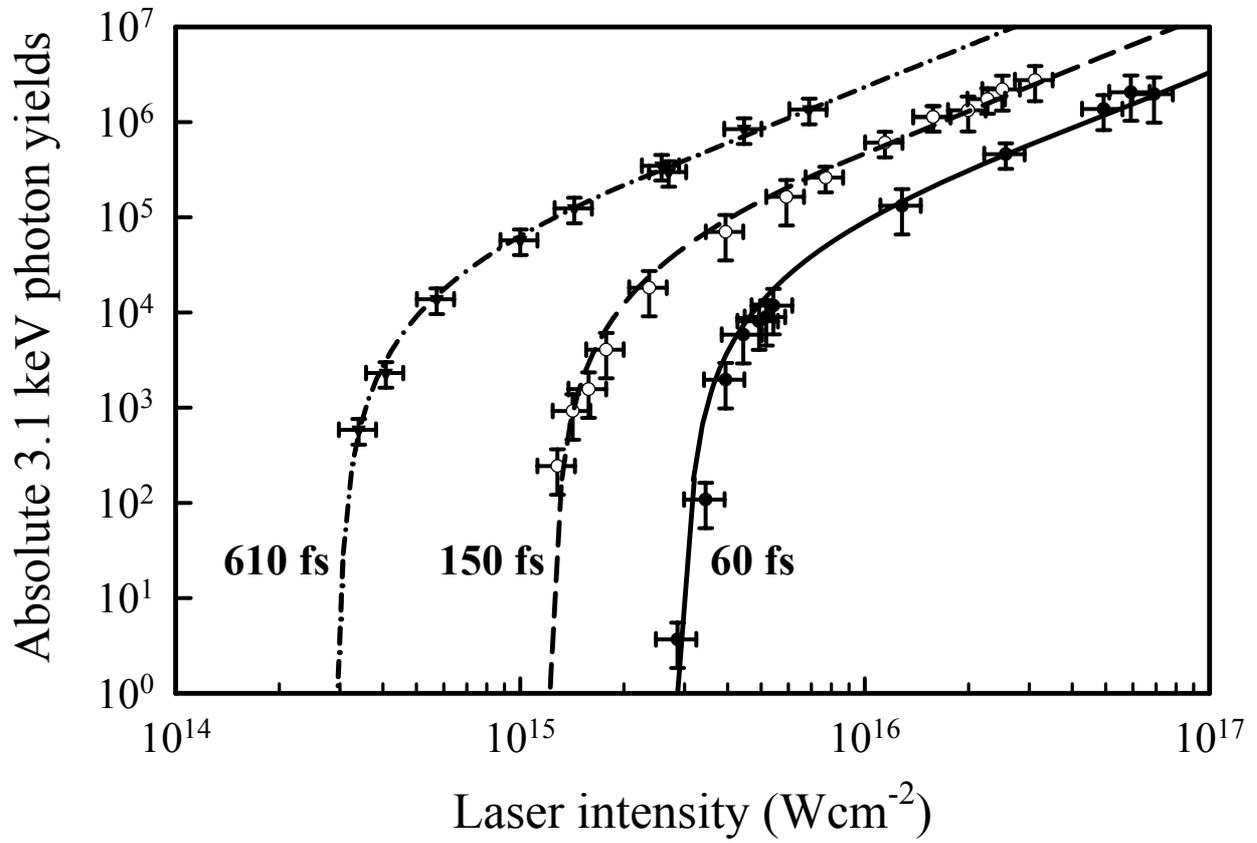

FIG. 2. X-ray yield variation with laser peak intensity obtained during irradiation of argon clusters by 60, 150 and 610 fs pulses at 800 nm with a backing pressure of 13 bars. Lines correspond to the calculated effective focal volume variation with laser intensity (see text).

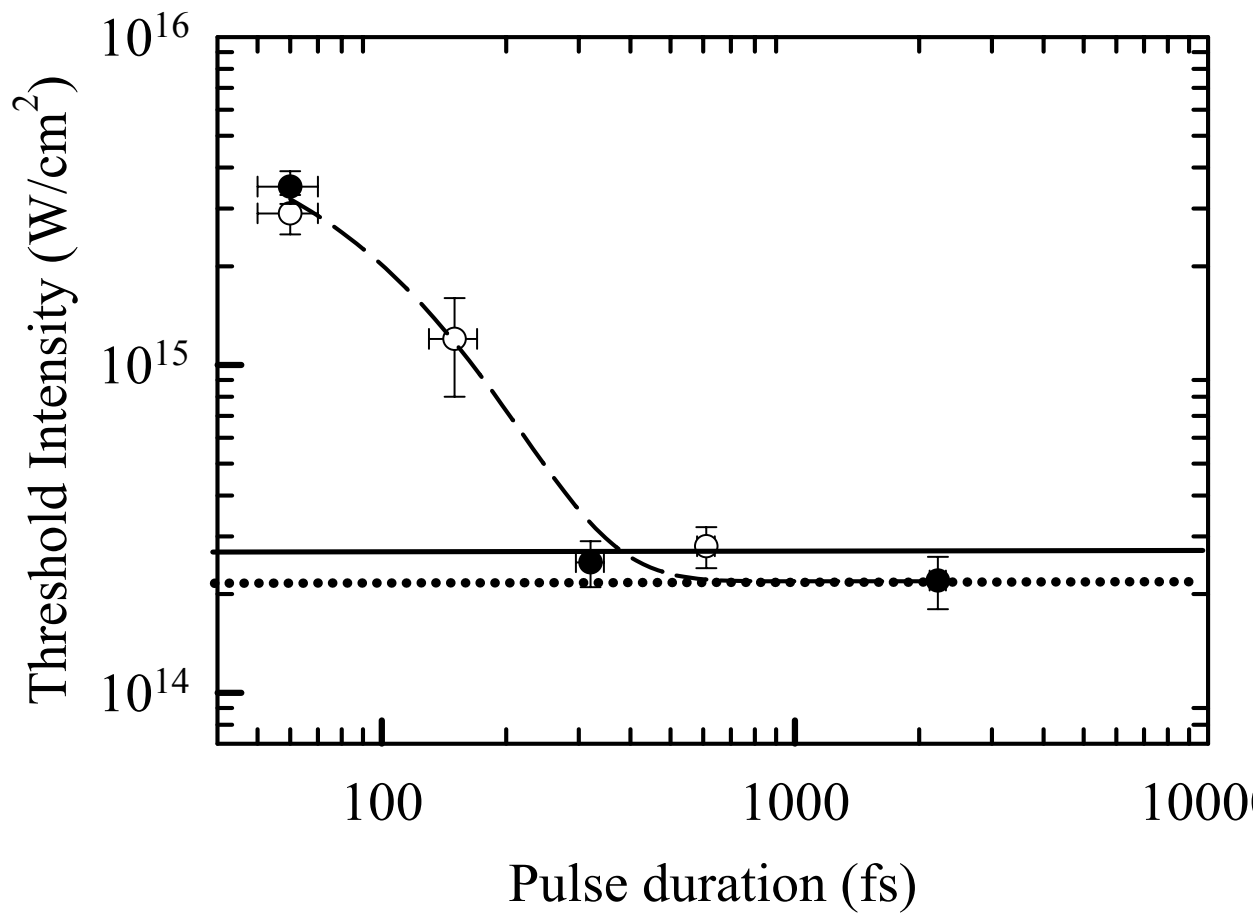

FIG. 3. Threshold intensities as a function of laser pulse duration for Ar and Xe clusters. Open (Ar) and closed (Xe) symbols are experimental results. Solid and dotted lines correspond to the OFI thresholds for producing $Ar^{1+}$ and $Xe^{2+}$ respectively. Dashed line is plotted to guide the eyes.

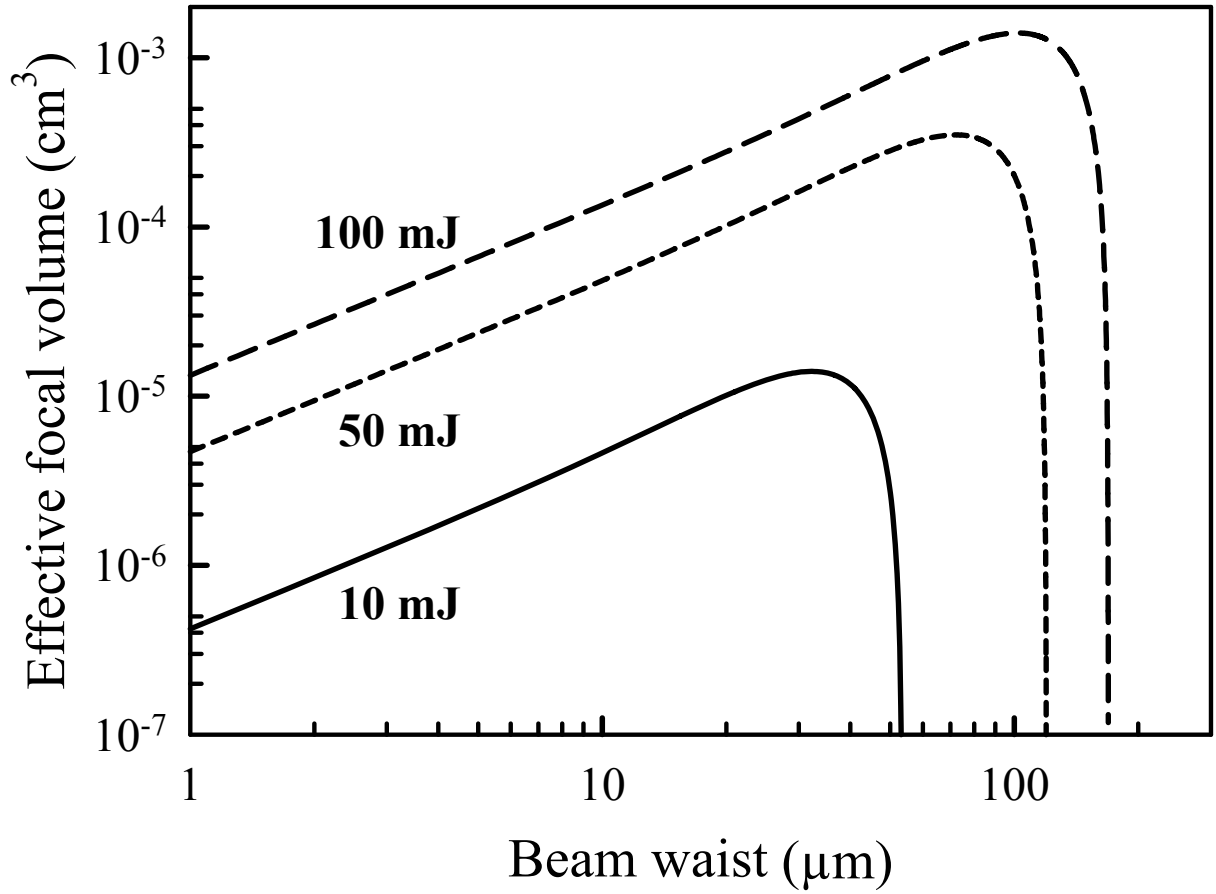

**Fig. 4.** Evolution of the effective focal volume with the beam waist for several laser energies. The threshold peak intensity and pulse duration are assumed to be $2 \times 10^{15}$ Wcm$^{-2}$ and 50 fs respectively.